\documentclass[aps,prb,twocolumn,showpacs,floatfix]{revtex4}
\usepackage{graphicx}
\usepackage{subfigure}
\usepackage{epsfig}
\newcommand{\bs}{\symbol{92}}

\begin{document}

\title{Light transmission assisted by Brewster-Zennek modes in chromium films carrying a subwavelength hole array}

\author{Micha\"{e}l Sarrazin}
\email{michael.sarrazin@fundp.ac.be} \affiliation{Laboratoire de
Physique du Solide, Facult\'es Universitaires Notre-Dame de la
Paix, \\rue de Bruxelles 61, B-5000 Namur, Belgium}

\author{Jean-Pol Vigneron}
\affiliation{Laboratoire de Physique du Solide, Facult\'es
Universitaires Notre-Dame de la Paix, \\rue de Bruxelles 61,
B-5000 Namur, Belgium}

\date{\today}

\begin{abstract}
This work confirms that not only surface plasmons but many other
kinds of electromagnetic eigenmodes should be considered in
explaining the values of the transmittivity through a slab bearing
a two-dimensional periodic corrugation. Specifically, the role of
Brewster-Zennek modes appearing in metallic films exhibiting
regions of weak positive dielectric constant. It is proposed that
these modes play a significant role in the light transmission in a
thin chromium film perforated with normal cylindrical holes, for
appropriate lattice parameters.
\end{abstract}

\pacs{78.20.-e, 42.79.Dj, 42.25.Bs, 42.25.Gy}

\maketitle

\section{Introduction}

The optical properties of thin metallic films containing periodic
arrays of subwavelength holes have been actively studied in
relation with the possibility of transmission engineering.
Originally, the remarkable optical transmission found in these
structures was pointed out by T.W. Ebbesen  \textit{et al} back in
1998\cite{Ebbesen-Nat-1998}. After these early experimental
observations, the role of the thin metallic film surface plasmons
(SPs) was put forward in order to explain the peculiar wavelength
dependence of the
transmission\cite{Ebbesen-Nat-1998,Schroter-1998,Thio-JOSA-1998,Ghaemi-PRB-1998,Pendry-PRL-1999,
Grupp-APL-2000,Thio-physica-2000,Krishnan-OC-2001,CAO-PRL-2002}.
As a consequence of this suggestion, it was generally believed
that the typical transmission features could only be obtained with
metallic films\cite{Ebbesen-Nat-1998,Grupp-APL-2000}. However it
is important to note that other explanations had been developed
which could compete with the SPs
model\cite{vigoureux-OC-2002,Treacy-PRB-2002,CAO-PRL-2002,Lezec-OE-2004,Sarrazin-PRB-2003,Genet-OC-2003}.
For instance, it was suggested that these phenomena could also be
described in terms of the short range diffraction of evanescent
waves\cite{vigoureux-OC-2002}, or in terms of dynamical
diffraction effects\cite{Treacy-PRB-2002}. Another explanation
suggests to emphasize the role of cavity resonances taking place
in the holes to explain the transmission
enhancement\cite{CAO-PRL-2002}.

\begin{figure}
\caption{A view of the system under study}\label{fig1}
\end{figure}

Many questions remain to completely clarify the scattering
processes involved in these experiments. Nevertheless, even if the
exact role of SPs is not clearly
assessed\cite{Ebbesen-Nat-1998,Schroter-1998,Thio-JOSA-1998,Ghaemi-PRB-1998,Pendry-PRL-1999,
Grupp-APL-2000,Thio-physica-2000,Krishnan-OC-2001,CAO-PRL-2002},
most interpretations tend to admit that SPs should play a key role
in the observed transmission spectrum features. The observed
transmission exhibits a set of peaks and dips. Many authors
suggested that transmission peaks are created by SPs
resonances\cite{Ebbesen-Nat-1998,Schroter-1998,Thio-JOSA-1998,Ghaemi-PRB-1998,Pendry-PRL-1999,
Grupp-APL-2000,Thio-physica-2000,Krishnan-OC-2001} but others
rather saw the SPs resonances at the spectral
dips\cite{CAO-PRL-2002}. In a recent work\cite{Sarrazin-PRB-2003}
we suggested to redefine the role of SPs in Ebbesen experiments in
the required context of resonant Wood
anomalies\cite{Wood-PR-1935}. In doing so, we have shown that the
transmission spectrum could be better depicted as a series of Fano
profiles\cite{Fano-AnnPhy-1938}. These recognizable lineshapes
result from the interference of non resonant transfers with
resonant transfers which involve the film SPs and evanescent
diffraction orders. We could then point out that each transmission
peak-dip pair is nothing else than a Fano
profile\cite{Fano-AnnPhy-1938}. The appearance of an assymetric
Fano lineshape does not necessarily locates the peak or the dip at
the SP resonance\cite{Sarrazin-PRB-2003}. However, we have shown
that the existence of the SPs is a condition for the presence
of the Fano lineshapes\cite{Sarrazin-PRB-2003}. It can be noted
that recent results by C. Genet \textit{et al} confirm this
description\cite{Genet-OC-2003}.

As a further outcome of this work\cite{Sarrazin-PRB-2003} it
became clear this kind of transmission spectrum, with its Fano
lineshapes, could also be obtained in a more general context and
that the generic concept of eigenmodes could substitute that of
SPs. These results implied two important new ideas. First,
according to the contributions of resonant and non-resonant
processes in the Fano profile, eigenmodes can be associated with
wavelengths closer to the peaks or to the dips. Second, it is
possible to obtain transmission curves similar to those of metal
films, by substituting SPs with guided modes or other types of
polaritons. Many examples can be found, including highly
refractive materials defining guided modes or ionic crystals in
the restrahlen band defining phonon-polaritons.

In a previous paper\cite{Sarrazin-PRB-2003}, it was made clear
that a strong hypothesis on the origin of the eigenmodes was not
required to account for the general features of the membrane
transmission.  As metal films were considered there, it seemed
natural to involve SPs in the transmission mechanisms. By
contrast, in another recent paper\cite{Sarrazin-PRE-2003},
simulations of light transmission through an array of
subwavelength cylindrical holes in a tungsten layer deposited on
glass substrate was given an interpretation based on dielectric
guided modes. Indeed, in the wavelength domain under examination,
tungsten exhibits a positive permittivity real part\cite{Palik91}.
Though SPs cannot exist, the transmission pattern is found very
similar to that obtained with a metallic film. These theoretical
results have recently been confirmed
experimentally\cite{Lezec-OE-2004}. Surprisingly, the very fact
that the typical transmission pattern can be observed even in
non-metallic systems convinced some authors\cite{Lezec-OE-2004} to
fully reject the SPs hypothesis and rather consider models
involving non-resonant evanescent waves diffraction. To our
knowledge, this point of view is not supported by recent
results\cite{Sarrazin-PRB-2003,Genet-OC-2003,Sarrazin-PRE-2003}
and the SPs must be replaced by other kinds of
eigenmodes\cite{Sarrazin-PRB-2003,Sarrazin-PRE-2003} in the
non-metallic cases.

The purpose of the present paper is to demonstrate that the
transmission profile of a chromium film, in the restricted
wavelength domain ($1112-1292$ $nm$) where the dielectric constant
is positive\cite{Palik91}, should involve eigenmodes which are not
SPs nor guided modes. In this case, SPs can be substituted by
Brewster-Zennek modes (BZ modes). This confirms that the
observation of the effects described by T.W. Ebbesen \textit{et
al} in systems where SPs do not exist do not preclude the
existence of a mechanism involving the excitation of other
eigenmodes.

\begin{figure}
\caption{Real ($\varepsilon _m^{\prime }$) and imaginary
($\varepsilon _m^{\prime \prime }$) part of the chromium
permittivity as a function of wavelength}\label{fig2}
\end{figure}

\section{Brewster-Zennek modes}

Let us recall some properties of surface modes, including the
concept of BZ modes\cite{Agarwal-PRB-1973,Yang-PRB-1991}. We
consider an interface between two media, $1$ and $2$, described by
their respective permittivities $\varepsilon _1$ and $\varepsilon
_2$. It is well know that for an isotropic material, only $p$
polarized modes can occur as a surface modes. We define $k$ as the
wave vector component parallel to the interface and we set $k_0 =
\omega /c$, where $\omega $ is the angular frequency of the mode.
Then, for the surface mode, the normal component of the wave
vector is given in each medium by
\begin{equation}\label{beta}
\beta _i=\sqrt{k^2-k_0^2\varepsilon _i}.
\end{equation}
The subscript $i$ denotes one of the media $1$ or $2$.

In the following we enforce a real positive value of $\varepsilon
_1$, which means that medium $1$ is a lossless dielectric
material, while we admit the form $\varepsilon _2=\varepsilon
_2^{\prime}+i\varepsilon _2^{\prime \prime }$ ($\varepsilon
_2^{\prime \prime}\geq 0$) for medium $2$, which means that this
medium is allowed to generate losses. Applying electromagnetic
boundary conditions pertinent to complex-response discontinuities,
we are lead to the following well-known expression which controls
the appearance of surface modes\cite{Raether88}
\begin{equation}\label{disp}
\varepsilon _1\beta _2+\varepsilon _2\beta _1=0
\end{equation}
From this, and eq. (\ref{beta}), we can develop the following
expression, out of which a dispersion relation can be drawn
\begin{equation}\label{k2}
k^2=k_0^2\frac{\varepsilon _1\varepsilon _2}{\varepsilon _1+\varepsilon _2}
\end{equation}
However, it is important to realize that in this expression, the
factors $\beta_{i}$ have been squared, so that eq. (\ref{k2}) and
(\ref{disp}) are not equivalent. The solutions of eq. (\ref{disp})
can be found among those of eq. (\ref{k2}), but some of the
solutions of eq. (\ref{k2}) are spurious and should be discarded.
One way to do this is to take the possible values of $k^{2}$ from
eq. (\ref{k2}), inject them in eq. (\ref{disp}), and check for the
correct matching. This actually generates a second condition,
which can be used together with eq. (\ref{k2}), to make precise
the acceptable eigenmodes branches. This condition turns out to be
\begin{equation}\label{cond}
\sqrt{\frac{-z^2}{1+z}}+z\sqrt{\frac{-1}{1+z}}=0
\end{equation}
where, somewhat surprisingly, the only variable that needs to be
examined is the ratio $z=\varepsilon_2/\varepsilon _1$. The
function of $z$, at the left-hand side of this equation, when
carefully calculated, defines a border line which splits the
complex plane into two extended connected regions. On one side of
this border, the left-hand side of eq. (\ref{cond}) cannot vanish,
while on the other side, it will vanish everywhere. The surface
states can only show up when the ratio $z$ lies in this latter
region.

The border separating the two regions can be obtained as follows.
Both terms added in eq. \ref{cond} are complex numbers with the
same modulus. In one of the regions mentioned above, they are
opposite complex numbers and they cancel each other; in the rest
of the complex plane, they turn out to be identical and they
cannot add to zero. The borderline will then be found at the
complex points $z$ where the sign determination of one of the
square roots in eq. \ref{cond} changes, and this requires to look
for the cut of the complex square-root function $\sqrt{Z}$. It is
convenient to take this cut as the straight segment $Z=-i\gamma$
($\gamma$ real and positive) and this choice lead us to identify
two distinct implicit paths,
\begin{equation}
-\frac 1{1+z}=-i\gamma
\end{equation}
and
\begin{equation}
-\frac{z^2}{1+z}=-i\gamma.
\end{equation}
which form the border. This choice of the cut is dictated in part
by the choice of a time-dependent oscillation written as $\exp
(-i\omega t)$ and the requirement that $\sqrt{1}=1$. Once the
borderline is known, a quick check at the case $z=1$ shows that no
cancellation can occur there, so that the active region where
eigenmodes appear is, unambiguously, that region which does not
contain $z=1$.

On the basis of this analysis, it is easily shown that, whatever
$\varepsilon_2=\varepsilon _2^{\prime }+i\varepsilon _2^{\prime
\prime }$ such that $\varepsilon _2^{\prime }<0$ and $\varepsilon
_2^{\prime \prime }\geq 0$ surface eigenmodes exist. These modes
are SPs and they will appear at metal/dielectric interfaces
regardless of the metal component of the heterojunction.

If $\varepsilon _2^{\prime }\geq 0$, surface eigenmodes exist only
when $0\leq \varepsilon _2^{\prime }<\varepsilon _1$ if the
condition
\begin{equation}
\varepsilon _2^{\prime \prime }>\varepsilon _2^{\prime }\sqrt{\frac{%
\varepsilon _1+\varepsilon _2^{\prime }}{\varepsilon _1-\varepsilon
_2^{\prime }}}
\end{equation}
is fulfilled.  Obviously such modes are not SPs and it is well
known that a dielectric/dielectric interface, without loss, cannot
support a surface mode if $\beta _i$ is real. Nevertheless,
surface modes can exist for a dielectric/dielectric interface if
one of the media is a lossy material such that the imaginary part
of its permittivity verify the above conditions. Such modes are
known as Brewster-Zenneck modes. They verify $\mathop{\rm Re}
\left\{ k\right\} <\frac \omega c\sqrt{\varepsilon _1}$ and are
necessarily related, by definition, to radiative surface modes
(contrasting SPs in metals such that iron, gold or copper).
Fundamentally it is well known that SPs consist in a collective
motion of electrons\cite{Raether88}. In the context of an
electromagnetic model, the electronic character of SPs is then
contained in the permittivity $\varepsilon_2$. Indeed, the
permittivity properties take their origin in the electronic
characteristics of the medium\cite{Raether88}. This is true
whatever the value of $\varepsilon _2^{\prime }$, positive or
negative. So, despite the above-mentioned differences between SPs
and BZ modes, both are related intrinsically to a collective
electronic phenomenon. In this way, the transition from the SPs
towards the BZ modes and \textit{vice versa} occurs without
discontinuities. Note that the present paper deals with metals
only. In the case of ionic crystals in the far infrared, it would
be natural to expect for the BZ modes counterpart associated with
phonon-polaritons instead of SPs.

\begin{figure}
\caption{(Color online) Comparison between transmission and
resonances. Solid line : $a=1000$ nm, Dashed line : $a=1200$ nm.
(a) : zeroth order transmission against wavelength. A and B are
the main maxima of transmission. (b) : Resonant diffraction orders
amplitudes related to the vectors $(0,\pm 1)$ of the reciprocal
lattice, at the substrate/metal interface (1) and vaccuum/metal
interface (2), as a function of the wavelength}\label{fig3}
\end{figure}

\section{Brewster-Zennek-based transmission}

Now we study the properties of the surface modes in the case of a
chromium grating similar to those submitted to experiments by Thio
{\it et al}\cite{Thio-JOSA-1998} (see fig. \ref{fig1}). In a large
part of the spectrum of interest here,  the real part of the
chromium permittivity is negative (fig. \ref{fig2}), as for simple
metals below the plasma frequency. However, for wavelengths in the
interval from $1112$ to $1292$ nm, it becomes positive, as,
specifically, we observe $0<\varepsilon _2^{\prime }<0,4$ and
$25,54<\varepsilon _2^{\prime \prime }<27,3$. With vacuum and
substrate (glass) permittivities equal to $1$ and $2,24$
respectively, the conditions for the formation of BZ modes are
verified at both interfaces of the chromium film. This is a
peculiar situation : with tungsten films in the dielectric
domain\cite{Palik91,Sarrazin-PRE-2003}, for instance, the same
conditions are not fulfilled and the BZ modes do not appear.

The transmission spectrum can easily be understood if we first
localize the BZ modes in the spectrum of a flat, homogeneous film.
For normal incident light falling on a such a homogeneous film
considered periodic on a square lattice of parameter $a$ (the \bs
empty'' lattice case), surface modes resonance will appear at the
wavelength $\lambda _{i,j}$ provided by the following
equation\cite{Raether88}
\begin{equation}\label{ij}
\lambda _{i,j}=\frac a{(i^2+j^2)^{1/2}} \mathop{\rm Re}
\sqrt{\frac{\varepsilon _1\varepsilon _2}{\varepsilon
_1+\varepsilon _2}}
\end{equation}
where $(i,j)$ denotes the related vector ${\bf g}$ of the reciprocal
lattice, such that
${\bf{g}} = \left( {{{2\pi } \mathord{\left/
 {\vphantom {{2\pi } a}} \right.
 \kern-\nulldelimiterspace} a}} \right)\,\left( {i,j} \right)$.
These wavelengths are indicative only, as they are based on a
model where the cylindrical holes have not been accounted for.

More detailed simulations which do not contain this simplification
have been carried out. The calculations which will now be
described are based on a coupled-modes method which combines a
scattering matrix formalism with a plane wave representation of
the fields. This technique provides a computation scheme for the
amplitude and polarization ($s$ or $p$) of reflected and
transmitted fields in any diffracted order. A brief account of the
method has been presented
elsewhere\cite{Sarrazin-PRB-2003,Sarrazin-OC-2004,
Vigneron-UM-1995} so that we will not recall any technicality
about these computations.

Following the experiments\cite{Thio-JOSA-1998}, fig. \ref{fig3}a
shows the calculated zeroth order transmission of a grating
prepared by perforating a $100$ nm thick chromium film deposited
on glass with circular holes of $250$ nm radius (fig. \ref{fig1}).
The two-dimensional square lattice parameter is $a=1000$ nm (solid
line) or $a=1200$ nm (dashed line). The incident beam is normal to
the grating surface and polarized so that the electromagnetic wave
electric field is oriented along one of the holes nearest-neighbor
directions ($Oy$ axis). A and B on Fig. \ref{fig3}a indicate the
main transmission maxima. Fig. \ref{fig3}b represents the
amplitude of the calculated resonant $p$-polarized diffraction
orders $(0,\pm 1)$ for the substrate/metal interface (peak 1) and
for the vacuum/metal interface (peak 2). Again, in Fig.
\ref{fig3}b, the dashed line refers to a lattice parameter of
$a=1200$ nm. It can be noted that, as justified by the above
discussion, these peaks do not correspond to the transmission
maxima. We have shown that the spectral lineshape of the resonance
can be interpreted as a Fano's
profile\cite{Sarrazin-PRB-2003,Genet-OC-2003}. What is remarkable
here is that, for a lattice parameter of  $a=1200$ nm, the
resonance has moved to a spectral region ($1200$ nm) where the
real part of the chromium permittivity is positive, and that
clearly SPs do not exist there. In this situation the resonance is
related to the appearance of a BZ mode. Fig \ref{fig4} shows the
location of both peaks (1) and (2) as a function of the grating
parameter $a$ (solid lines (1) and (2), respectively). We could
compare these values with the wavelengths obtained from eq.
\ref{ij} (dashed lines (1) and (2) respectively) for the
appropriate diffraction orders. A good agreement is found between
these calculated quantities. We show that as the lattice parameter
increases, the resonance wavelength increases linearly as
predicted by eq. \ref{ij}. The hatched domain in fig. \ref{fig4}
is characterized by the presence of BZ modes. The SP domain lies
everywhere else. One interesting observation is that both kinds of
surface modes (BZ modes and SPs modes) substitute each others in a
continuous way in accordance with our previous explanation in
section II. The location of transmission maxima A (circle dots)
and B (square dots) are also shown on fig. \ref{fig4}. These, as
already underlined, do not indicate directly the eigenmodes
resonance wavelengths though these transmission maxima are known
to be related to the surface modes via a Fano profile.

\begin{figure}
\caption{Position of both peaks (1) and (2) against grating
parameters $a$. Solid lines (1) and (2) respectively are obtained
from numerical computations. Dashed lines (1) and (2) are obtained
from eq.(8). A (circle dots) and B (square dots) : location of
both main maxima of transmission against $a$. Hatched domain : BZ
modes, elsewhere : SPs modes.}\label{fig4}
\end{figure}

Note that, in a previous work\cite{Ghaemi-PRB-1998}, it was
inferred that the SPs wavelength were red-shifted, compared to
those calculated for an empty lattice. It was assumed that such a
shift was needed to make the transmission maxima coincide with SPs
wavelengths and the justification suggested for this shift was
based on the presence of holes. This is not observed in actual
detailed calculations. The present results show that the exact
resonant wavelengths of the surface modes, as computed
numerically, are not shifted enough to justify their location on
transmission maxima. Nevertheless, as mentioned before, an
eigenmode resonance does not necessarily coincide with a transmission
maximum, but a maximum-minimum pair indicates the presence of a nearby eigenmode
resonance\cite{Sarrazin-PRB-2003}. In fact, in this case, the surface modes
resonances are incidentally located closer to the minima of the
transmission. It is interesting to note that Thio \textit{et al} have
compared experimentally the transmission from gratings similar to ours, with lattice parameters $a=1000$ nm and $a=1200$ nm\cite{Thio-JOSA-1998}. The present computations of the transmission matched very well
with these data\cite{Thio-JOSA-1998}. We take this as a strong
evidence of the relevance of the Brewster-Zennek modes in the description of these experiments.

\section{Conclusion}

In the context of studies of the optical transmission properties
of subwavelength hole arrays, we have shown that Brewster-Zennek
modes can substitude  surface plasmons when considering a layer
made from a lossy dielectric medium, for specific lattice
parameters. When modelling a grating etched in a chromium film we
can use the concept of Brewster-Zennek modes to explain the
spectra found in previous experiments. This study recalls the
existence requirement of eigenmodes for observing the typical
transmission lineshapes and we have shown that, contrasting a
frequent believe, surface modes wavelengths based on an empty
lattice model are not significatively red-shifted to justify their
location on transmission maxima.

\begin{acknowledgments}
We acknowledge the use of the Namur Interuniversity Scientific
Computing Facility (Namur-ISCF), a joint project between the
Belgian National Fund for Scientific Research (FNRS), and the
Facult\'{e}s Universitaires Notre-Dame de la Paix (FUNDP).

This work was partially supported by the EU Belgian-French
INTERREG III project \bs PREMIO'', the EU5 Centre of Excellence
ICAI-CT-2000-70029 and the Inter-University Attraction Pole (IUAP
P5/1) on \bs Quantum-size effects in nanostructured materials'' of
the Belgian Office for Scientific, Technical, and Cultural
Affairs.
\end{acknowledgments}

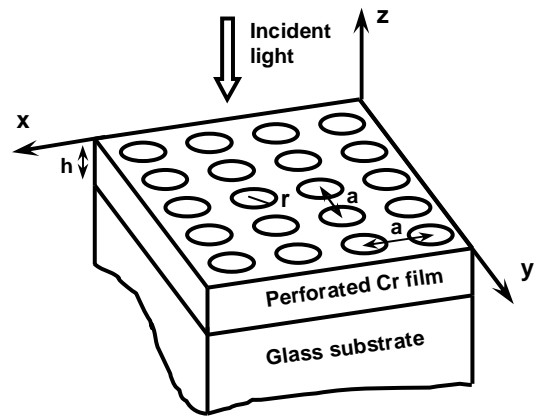
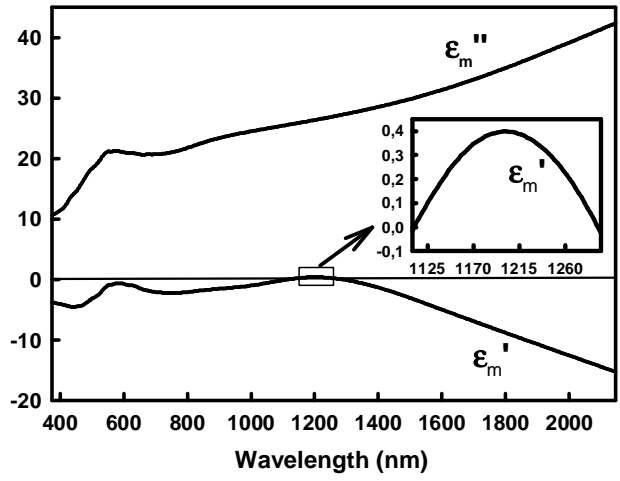
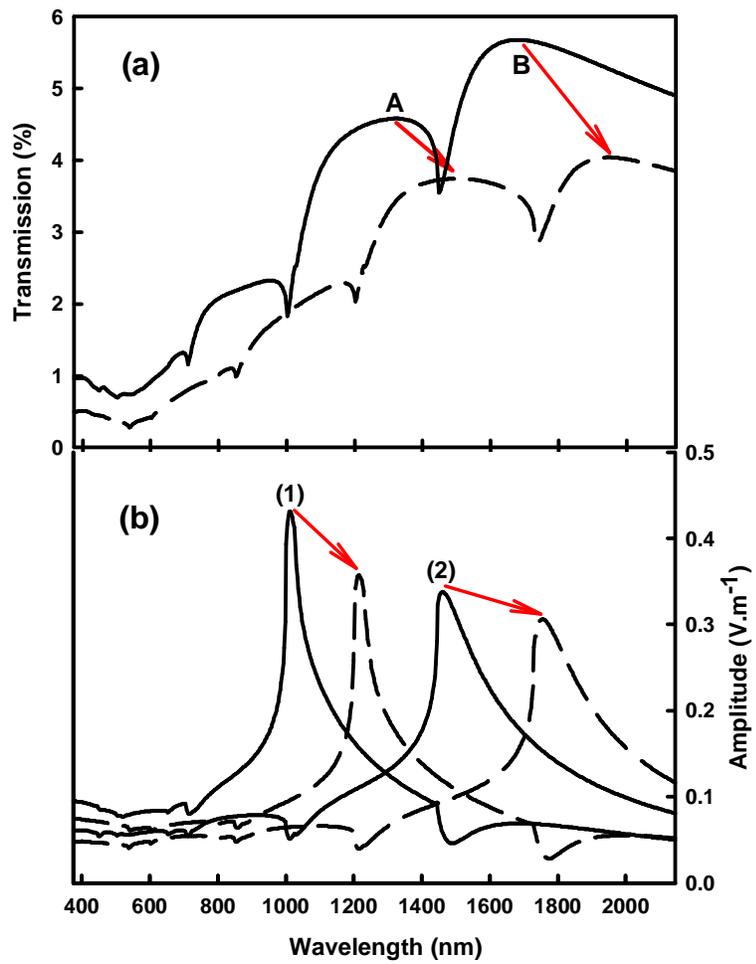
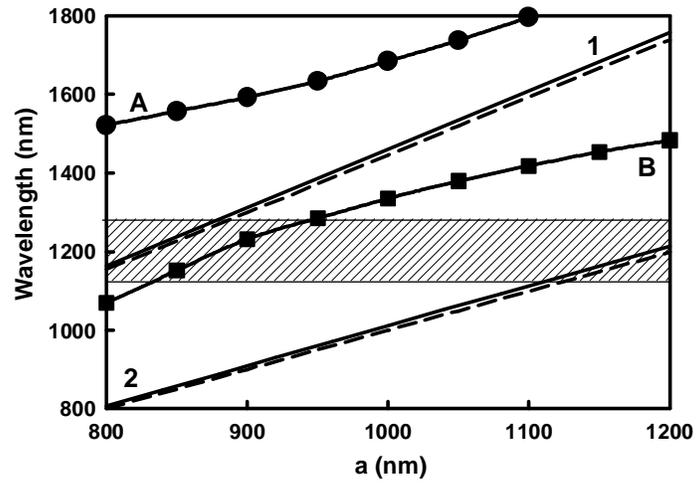
\end{document}